\global\def\draftcontrol{0}

   \def\versionno{refinement of Seiberg-Witten}

\catcode`\@=11

\expandafter\ifx\csname draftcontrol\endcsname\relax\global\def\draftcontrol{0} 
\fi 

{\count255=\time\divide\count255 by 60 
\xdef\hourmin{\number\count255} 
\multiply\count255 by-60\advance\count255 by\time 
\xdef\hourmin{\hourmin:\ifnum\count255<10 0\fi\the\count255}} 
\def\draftdate{\number\month/\number\day/\number\year\ \ \ \hourmin } 

\newcommand\makepapertitle{\par

  \begingroup 
    \renewcommand\thefootnote{\@fnsymbol\c@footnote}%
    \def\@makefnmark{\rlap{\@textsuperscript{\normalfont\@thefnmark}}}%
    \long\def\@makefntext##1{\parindent 1em\noindent 
            \hb@xt@1.8em{%
                \hss\@textsuperscript{\normalfont\@thefnmark}}##1}%
     \newpage 
     \global\@topnum\z@   
     \@makepapertitle 
     \thispagestyle{empty}\@thanks 
  \endgroup 
  \setcounter{footnote}{0}%
  \global\let\thanks\relax 
  \global\let\makepapertitle\relax 
  \global\let\@makepapertitle\relax 
  \global\let\@thanks\@empty 
  \global\let\@author\@empty 
  \global\let\@date\@empty 
  \global\let\@title\@empty 
  \global\let\title\relax 
  \global\let\author\relax 
  \global\let\date\relax 
  \global\let\and\relax 
  \def\version{\let\version\@version\@gobble} 
} 
\def\@makepapertitle{%
  \newpage 
   \ifnum\draftcontrol=1 {} 
   \version\versionno 
   \vskip 5.5em%
   \else 
   \hfill\hbox to 3cm {\parbox{4.5cm}{\@pubnum}\hss}%
   \vskip 6.5em%
   \fi 
   \begin{center}%
   \let \footnote \thanks 
      {\hskip -0\textwidth \hbox to 1\textwidth%
        {\centerline{\Large\bf{\noindent\@title}}}}%
     \vskip 2em%
     {\normalsize
       \lineskip .5em%
       \begin{tabular}[t]{c}%
         \@author 
       \end{tabular}\par}%
     \vskip 1.5em%
     {\@bstract}%
     \end{center}%
     \vfill
     \@date%
     \vskip 1.5em%
   \par 
} 

\gdef\@pubnum{} 
\def\pubnum#1{%
  \gdef\@pubnum{#1}} 

\gdef\@bstract{} 
\def\Abstract#1{%
  \gdef\@bstract{%
   \parbox{\textwidth-0pc}{%
   \centerline{\bf Abstract}\penalty1000 
   \noindent
   \renewcommand\baselinestretch{1.0} 
   {#1}}} 
} 

\gdef\@email{}
\def\email#1{%
   \gdef\@email{%
   Email: {\tt #1}}
}

\def\ps@paper{\let\@mkboth\@gobbletwo%
     \ifnum\draftcontrol=1 
        \def\@oddfoot{\hbox to \textwidth{\tiny \versionno \hfil\tiny\draftdate}%
        \hskip -\textwidth \hbox to \textwidth{\hfil\rm\thepage\hfil}}%
     \else\def\@oddfoot{\hbox to \textwidth{\hfil\rm\thepage\hfil}} 
     \fi 
     \let\@evenfoot\@oddfoot 
} 

\def\body{\clearpage 
          \pagestyle{paper} 
        } 
\newenvironment{acknowledgments}{%
\vskip 3.25ex 
\addcontentsline{toc}{section}{Acknowledgments}
\noindent {\bf Acknowledgments} 
} 

\def\@version#1{\ifnum\draftcontrol=1 
\typeout{}\typeout{#1}\typeout{} 
\vskip3mm\centerline{\hbox{\fbox{\normalsize{\tt DRAFT -- #1 -- } 
                   {\draftdate}}}}\vskip3mm 
\fi} 
\let\version\@version 
\long\def\eqlabel#1{\ifnum\draftcontrol=1 
                    \tag@false  
                    \tag*{(\theequation) \hbox to -0.2cm{\hspace{0cm}\small{#1}\hss}} 
                    \refstepcounter{equation}  
                    \edef\@currentlabel{\theequation} 
                    \ltx@label{#1}          
                    \else 
                    \label{#1} 
                    \fi 
                    } 
\let\st@bibitem\@bibitem 
\let\st@lbibitem\@lbibitem 
\ifnum\draftcontrol=1 
  \def\@bibitem#1{%
    \st@bibitem{#1}\a@@label{#1}\ignorespaces} 
  \def\@lbibitem[#1]#2{%
    \st@lbibitem[#1]{#2}\a@@label{#2}\ignorespaces} 
  \def\a@@label#1{%
    \gdef\a@lab{\smash{\normalfont\small#1}} 
    \ifvmode 
      \if@inlabel 
        \global\setbox\@labels\hbox{%
          \llap{\a@lab\let\a@lab\relax 
                \kern\@totalleftmargin\kern\marginparsep}%
          \box\@labels}%
      \fi 
    \fi} 
\fi 

\documentclass[12pt,letterpaper]{article} 

\usepackage{amsmath,bm,amsfonts,amssymb,array,calc,amsthm,rotating}
\usepackage{epsfig,psfrag} 
\usepackage{rotating}
\usepackage{amscd}
\usepackage{graphicx}
\usepackage{color}
\usepackage[colorlinks=false]{hyperref}

\renewcommand\baselinestretch{1.25} 
\renewcommand\section{\@startsection {section}{1}{\z@}%
                                   {-3.5ex \@plus -1ex \@minus -.2ex}%
                                   {2.3ex \@plus.2ex}%
                                   {\normalfont\large\bfseries}} 
\renewcommand\subsection{\@startsection{subsection}{2}{\z@}%
                                   {-3.25ex\@plus -1ex \@minus -.2ex}%
                                   {1.5ex \@plus .2ex}%
                                   {\normalfont\normalsize\bfseries}} 
\renewcommand\subsubsection{\@startsection{subsubsection}{3}{\z@}%
                                   {-3.25ex\@plus -1ex \@minus -.2ex}%
                                   {1.5ex \@plus .2ex}%
                                   {\normalfont\normalsize\it}} 
\renewcommand\paragraph{\@startsection{paragraph}{4}{\z@}%
                                   {-1.75ex\@plus -1ex \@minus -.2ex}%
                                   {1ex \@plus .2ex}%
                                   {\normalfont\normalsize\bf}} 
\renewcommand\subparagraph{\@startsection{subparagraph}{5}{\z@}%
                                   {-1.25ex\@plus -0ex \@minus -.2ex}%
                                   {-2ex \@plus .2ex}%
                                   {\normalfont\normalsize\it}}

\numberwithin{equation}{section}

\long\def\@makecaption#1#2{%
  \vskip\abovecaptionskip
  \sbox\@tempboxa{{\bf #1:} #2}%
  \ifdim \wd\@tempboxa >\hsize
    {\small\bf #1:} {\small #2}\par
  \else
    \global \@minipagefalse
    \hb@xt@\hsize{\hfil\box\@tempboxa\hfil}%
  \fi
  \vskip\belowcaptionskip}

\renewcommand*\l@section[2]{%
  \ifnum \c@tocdepth >\z@
    \addpenalty\@secpenalty
    \addvspace{.5em \@plus\p@}%
    \setlength\@tempdima{1.5em}%
    \begingroup
      \parindent \z@ \rightskip \@pnumwidth
      \parfillskip -\@pnumwidth
      \leavevmode \bfseries
      \advance\leftskip\@tempdima
      \hskip -\leftskip
      #1\nobreak\hfil \nobreak\hb@xt@\@pnumwidth{\hss #2}\par
    \endgroup
  \fi}
\renewcommand*\l@subsection{\addvspace{.0em \@plus\p@}\@dottedtocline{2}{1.5em}{2.3em}}
\renewcommand*\l@subsubsection{\addvspace{-.2em \@plus\p@}\@dottedtocline{3}{3.8em}{3.2em}}

\def\hepth#1{\href{http://xxx.arxiv.org/abs/hep-th/#1}{{arXiv:hep-th/#1}}}

\def\math#1{\href{http://xxx.arxiv.org/abs/math/#1}{{arXiv:math/#1}}}

\def\mathag#1{\href{http://xxx.arxiv.org/abs/math.AG/#1}{{arXiv:math.ag/#1}}}

\def\arxiv#1#2{\href{http://xxx.arxiv.org/abs/#1}{{arXiv:#1 [#2]}}}

\definecolor{refcol}{rgb}{0.2,0.2,0.8}
\definecolor{eqcol}{rgb}{.6,0,0}
\definecolor{purple}{cmyk}{0,1,0,0}

\gdef\@citecolor{refcol}
\gdef\@linkcolor{eqcol}
\def\colorlinkspurple{\gdef\@urlcolor{purple}}
\def\colorlinksblue{\gdef\@urlcolor{blue}}
\def\colorlinksred{\gdef\@urlcolor{red}}

\def\ie{{\it i.e.}} 
\def\eg{{\it e.g.}} 
\def\etc{{\it etc.}}

\def\revise#1       {\raisebox{-0em}{\rule{3pt}{1em}}%
                     \marginpar{\raisebox{.5em}{\vrule width3pt\ 
                     \vrule width0pt height 0pt depth0.5em 
                     \hbox to 0cm{\hspace{0cm}{%
                     \parbox[t]{4em}{\raggedright\footnotesize{#1}}}\hss}}}}

\def\calc         {{\cal C}}

\def\calf         {{\cal F}} 
\def\calg         {{\cal G}}

\def\calk         {{\cal K}} 
\def\call         {{\cal L}} 
\def\calm         {{\cal M}} 
\def\caln         {{\cal N}} 
\def\calo         {{\cal O}}

\def\calt         {{\cal T}}

\def\calw         {{\cal W}}

\def\complex      {{\mathbb C}} 
 
\def\projective   {{\mathbb P}} 
 
\def\reals        {{\mathbb R}} 
\def\zet          {{\mathbb Z}}

\def\del          {\partial} 
\def\delbar       {\bar\partial} 
\def\ee           {{\it e}} 
\def\ii           {{\it i}}

\newcommand\topa[2]{\genfrac{}{}{0pt}{2}{\scriptstyle #1}{\scriptstyle #2}}

\def\sqr#1#2{{\vcenter{\vbox{\hrule height.#2pt   
 \hbox{\vrule width.#2pt height#1pt \kern#1pt 
 \vrule width.#2pt}\hrule height.#2pt}}}}

\newcommand{\beq}{\begin{equation}}
\newcommand{\eq}{\end{equation}}
\newcommand{\req}[1]{(\ref{#1})}
\renewcommand{\ie}{{\it i.e.}}

\newcommand{\Fcal}{\mathcal F}

\newcommand{\Mcal}{\mathcal M}

\newcommand{\Tcal}{\mathcal T}

\newcommand{\Z}{\mathbb Z}
\newcommand{\R}{\mathbb R}
\newcommand{\T}{\mathbb T}
\newcommand{\C}{\mathbb C}

\newcommand{\m}{m}

\newcommand{\epo}{\epsilon_1}
\newcommand{\ept}{\epsilon_2}

\catcode`\@=12 

\begin{document} 


\title{Extended Holomorphic Anomaly in Gauge Theory}

\pubnum{
CERN-PH-TH-2010-149 \\
IPMU10-0108
}
\date{July 2010}

\author{
Daniel Krefl$^{a}$ and Johannes Walcher$^{b}$ \\[0.2cm]
\it ${}^{a}$ IPMU, The University of Tokyo, Kashiwa, Japan \\
\it $^{b}$ PH-TH Division, CERN, Geneva, Switzerland
}

\Abstract{
The partition function of an $\caln=2$ gauge theory in the $\Omega$-background satisfies,
for generic value of the parameter $\beta=-{\epo}/{\ept}$, the, in general extended, 
but otherwise $\beta$-independent, holomorphic anomaly equation of special geometry. Modularity 
together with the ($\beta$-dependent) gap structure at the various singular loci in the moduli space
completely fixes the holomorphic ambiguity, also when the extension is non-trivial. 
In some cases, the theory at the orbifold radius, corresponding to $\beta=2$, can be identified 
with an ``orientifold'' of the theory at $\beta=1$. The various connections give hints for 
embedding the structure into the topological string.
}

\makepapertitle

\body

\version\versionno

\vskip 1em



\section{Introduction}

Four-dimensional $\caln=2$ supersymmetric gauge theory was solved following Seiberg and Witten 
\cite{sw1,sw2} by exploiting constraints from special geometry and modular invariance on the moduli 
space of vacua. Subsequently, the theory and its solution were related to string theory in 
several ways, engendering a variety of developments that have revolutionized our understanding 
of the dynamics of theories with 8 supercharges.

A comparably early line of investigation was the regularization of the integral over the moduli 
space of instantons proposed in \cite{lns1,mns1,lns2,mns2}. It culminated in the verification of 
the Seiberg-Witten prepotential directly from the instanton counting \cite{nekrasov}. As a result,
the central object today is the partition function of the $\caln=2$ supersymmetric gauge theory
in the so-called $\Omega$-background,
\begin{equation}
\eqlabel{central}
Z(a,\epo,\ept;q) \,.
\end{equation}
Here, $a$ are the vectormultiplet moduli, $\epo, \ept$ are the equivariant parameters for the 
localization with respect to the two-dimensional torus acting on $\reals^4\cong\complex^2$, 
$q$ is the instanton counting parameter (related to the dynamical scale of the gauge theory, 
if any), and we have left other parameters such as masses of any matter fields implicit.

It was shown in \cite{nekrasov,neok}, see also \cite{nayo1}, that 
\begin{equation}
\eqlabel{limit}
\lim_{\epo,\ept\to 0} \Bigl(\epo\ept\log Z(a,\epo,\ept;q)\Bigr) = \calf^{(0)}(a;q)
\end{equation}
reproduces the prepotential computed by Seiberg and Witten from the periods of a family of 
complex curves. Following \cite{mowe,lns1,nekrasov}, the terms of higher order in $\epo,\ept$ 
in \eqref{limit} have been expected to capture some gravitational couplings of the gauge theory
arising from the embedding into string theory. This has been made precise for the terms of 
second order in $\epo,\ept$ in \cite{mowe,lns1}, but is straightforward in higher order only for
$\epo=-\ept$ \cite{nekrasov}. In that case, one is talking about terms of the form
$\int d^4\theta \Fcal^{(g)}\calw^{2g}$,
where $\calw$ is the self-dual gravi-photon chiral field, and $\calf^{(g)}$ can be obtained
as the field theory limit of the genus-$g$ topological string amplitude \cite{agnt,bcov} on 
the appropriate Calabi-Yau background constructed in geometric engineering \cite{klmvw,kkv}.

The computation of the higher order corrections to \eqref{limit} from the topological 
string perspective has been pursued for example in \cite{kmt,iqka,egka}. One of the advantages 
is that while the field theory localization technique is applicable only in the weak-coupling 
regime, the topological string machinery can yield results that are valid also in expanding 
around other interesting points in the moduli space. This is achieved in a familiar way by 
the method of the holomorphic anomaly that trades holomorphicity for modular invariance 
\cite{bcov}.

To be fair, of course, the field theory limit commutes with the holomorphic anomaly, and 
one can study the $\calf^{(g)}$ using just the input from special geometry provided by 
the Seiberg-Witten curve. This reasoning was followed in \cite{hukl1,hukl2}. In fact, 
it was found in these works that modularity together with the gap structure in the 
expansion of the amplitudes around the monopole/dyon singularities provides 
enough boundary conditions to completely fix the ambiguity that plagues the 
method of the holomorphic anomaly \cite{ghva}. 

These investigations have been restricted to the special self-dual background 
$\epo=-\ept$. The purpose of the present note is to shed light on $Z(a,\epo,\ept;q)$
from the perspective of the topological B-model, Seiberg-Witten geometry, special
geometry, and the holomorphic anomaly, but for {\it general values of $\epo,\ept$}.
A priori, it is not clear that the relation will persist, or what form it will take.
In particular, the topological string is not known to admit a two-parameter expansion
corresponding to $\epo,\ept$ (but see \cite{ahnt} for a recent proposal). Our results
are surprisingly simple: The holomorphic anomaly equation merely experiences a {\it slight 
extension} \cite{extended} by some data contained in \eqref{limit} to the {\it first order} 
in $\epo,\ept$, and the essential modification of the formalism is at the level of 
{\it fixing the boundary conditions}, which we are also able to determine completely.
We will also find a surprising relation between the theory at the special value 
$\epo=-2\ept$ and a certain ``orientifold'' (descending from the real topological 
string \cite{bouchard1,tadpole,real}) of the theory at $\ept=-\epo$. We view our results
mostly as an encouraging proof of principle that mirror symmetry continues
to make sense for general $\epo,\ept$. 

We will study in this paper $SU(2)\subset U(2)$ gauge theory coupled to $N_f=0,1,2,3$ 
massless hypermultiplets in the fundamental representation. We are confident that the 
structures we find carry over to other cases. We will proceed in the next section via 
special geometry, the holomorphic anomaly, the holomorphic limits and singularity structure 
at interesting points in moduli space. We then describe the four examples, each of which 
has some special illuminating features. We present the interpretation of the extension 
of the holomorphic anomaly equation from the point of view of Seiberg-Witten geometry in 
section \ref{more}. The orientifold relation can be found in section \ref{orientifold}.
Many readers will be familiar with most of the formulas, so we have relegated a lot
of technical baggage to the appendix, and the references.

\section{The Expansion}
\label{develop}

Motivated in part by recent developments relating $\caln=2$ gauge theory with two-dimensional
conformal field theory and matrix models \cite{agt,diva}, we begin by reparameterizing the 
$\Omega$-background according to
\begin{equation}
\eqlabel{redef}
\epo= \lambda \beta^{1/2} \,,\qquad  \ept=-\lambda \beta^{-1/2} \,.
\end{equation}
For small $q$ and fixed $\beta$, we then expand \eqref{central} in $\lambda$, 
which one might think of as 
the topological string coupling constant $g_s$ (or the Planck constant $\hbar$),
\footnote{The index $n$ is best thought of as running over the possible Euler numbers
of Klein surfaces, a point to which we shall return.}
\begin{equation}
\eqlabel{extracted} 
\log Z(a,\epo,\ept;q) \sim \sum_{n=-2}^\infty \lambda^n \calg^{(n)}(a,\beta;q) \,.
\end{equation}
According to \eqref{limit} we have
\begin{equation}
\eqlabel{prepotential}
\calg^{(-2)} = \calf^{(0)} \,,
\end{equation}
the Seiberg-Witten prepotential. In particular, it is $\beta$-independent. 
The term at order $\lambda^{-1}$ will play a central role in our story. It takes the
form
\begin{equation}
\eqlabel{extension}
\calg^{(-1)} = \bigl(\beta^{1/2}-\beta^{-1/2}\bigr) \calt \,,
\end{equation}
with $\calt$ independent of $\beta$. Notice that $\calg^{(-1)}$ vanishes in the standard topological 
string limit $\beta=1$.

To proceed, we review the role of $\calf^{(0)}$ in special geometry. We denote by $u$ a global 
coordinate on the moduli space $\calm$ of vacua, which is identified with the base space of an 
appropriate family of complex curves, $\calc_u$.\footnote{Many formulas we write will be restricted 
to a one-dimensional moduli space, or $SU(2)$ gauge theory with fundamental flavors. Generalizations 
are mostly obvious.} The family of curves is equipped with a meromorphic one-form 
$\lambda_{\rm SW}$, such that for approriate choice of one-cycles $A$ and $A_D$,
\begin{equation}
\eqlabel{periods}
a = \oint_A \lambda_{\rm SW} \,,\qquad a_D = \oint_{A_D} \lambda_{\rm SW}\,, 
\end{equation}
and
\begin{equation}
a_D = \frac{\del\calf^{(0)}}{\del a} \,,
\end{equation}
after eliminating $u$ from \eqref{periods}. We will not need to be explicit about the auxiliary
geometric data until later. For expansion in different regions of moduli space, it is anyways more 
convenient to base the developments on the Picard-Fuchs differential equation, a third order system 
of linear differential equations,
\begin{equation}
\eqlabel{PFeqs}
\call \varpi(u) = 0 \,,
\end{equation}
satisfed by all periods of $\lambda_{\rm SW}$. Using $a$ as a local coordinate around 
$u\to\infty$, the Picard-Fuchs operator takes the form\footnote{As is now evident, the
constant is a third solution of the differential equation. This solution decouples in special
cases, such as $SU(2)$ gauge theory with massless hypermultiplets.}
\begin{equation}
\eqlabel{flatPF}
\call = \del_a\frac{1}{C_{aaa}} \del_a^2 \,,
\end{equation}
where
\begin{equation}
C_{aaa} = \del_a^3 \calf^{(0)} = \del^2_a a_D(a) = \del_a \tau(a)
\end{equation}
is a (meromorphic) rank three symmetric tensor over $\calm$, referred to as the Yukawa 
coupling, that plays a central role in special geometry. One feature of special geometry is
the existence of canonical (flat) coordinates \cite{bcov}, providing a meaningful expansion 
parameter around any interesting point $u=u_*$ in $\calm$. In such a flat coordinate 
$t=t(u)$, vanishing 
at $u=u_*$, the Picard-Fuchs operator takes again the form \eqref{flatPF} with $a\to t$, \ie,
\begin{equation}
\eqlabel{general}
\call=\del_t\frac{1}{C_{ttt}}\del_t^2\,,\qquad
C_{ttt} = \Bigl(\frac{\del u}{\del t}\Bigr)^3 C_{uuu} \,.
\end{equation}
A useful property of the canonical coordinates is that in the holomorphic limit 
$\bar{t}\to 0$ (or $\bar a\to\infty$ for $t=a$), the connection of the Weil-Petersson metric 
$g\sim {\rm Im}\tau$ on $\calm$ takes the form
\begin{equation}
\eqlabel{relation}
\lim_{\bar t\to 0} \Gamma^{u}_{uu} = \del_u \log \frac{\del t(u)}{\del u} \,.
\end{equation}

At this stage, we are ready to write down the holomorphic anomaly equations of \cite{bcov}.
According to \cite{bcov}, the amplitudes $\calf^{(g)}(a)$ extracted from \eqref{extracted}
via
\begin{equation}
\calf^{(g)}(a;q)=\calg^{(2g-2)}(a,\beta = 1;q) \,,
\end{equation}
while holomorphic in $a$, are not well-behaved globally over $\calm$. Instead, one should view the 
$\calf^{(g)}(a)$ (for $g\ge 1$) as the holomorphic limit $\bar a\to\infty$ of {\it non-holomorphic, 
but globally defined} objects $\calf^{(g)}(u,\bar u)$. (These are denoted by the same letter, as 
confusion can not arise.) 

For $g>1$, the amplitudes $\calf^{(g)}(u,\bar u)$ satisfy the holomorphic anomaly equation
\begin{equation}
\eqlabel{hae}
\delbar_{\bar u} \calf^{(g)} = \frac 12  \sum_{\topa{g_1+g_2=g}{g_i>0}}
{\bar C}_{\bar u}^{\; u u}\calf^{(g_1)}_u \calf^{(g_2)}_u  + 
\frac12{\bar C}_{\bar u}^{\; u u} \calf^{(g-1)}_{uu} \,,
\end{equation}
where $\calf^{(g)}_{uu}=D_u \calf^{(g)}_u = D_u^2 \calf^{(g)}$, $D_u$ is the covariant 
derivative over $\calm$, and indices are raised and lowered using the Weil-Petersson metric.
The one-loop amplitude satisfies the special equation
\begin{equation}
\eqlabel{oneloop}
\delbar_{\bar u}\del_u \calf^{(1)} =  \frac 12 {\bar C}_{\bar u}^{\; uu} C_{uuu} \,.
\end{equation}

It is a natural question to ask how \eqref{hae} should be modified away from $\beta=1$.
The structure of the expansion of \eqref{central} (see appendix), and in particular the 
generic non-vanishing of the terms of odd order in $\lambda$, suggests a possible role 
for the extended holomorphic anomaly equation of \cite{extended}. In the orientifold context 
of \cite{tadpole}, the amplitudes $\calg^{(n)}$ are the sums of all contributions at
fixed order in string perturbation theory. When $n$ is odd, these arise only from open
and unoriented diagrams, while when $n$ is even, we have $\calg^{(n)} \sim\calf^{(n/2+1)} + 
\cdots$. (Note that in the context of topological string orientifolds, the $\calg^{(n)}$
do not depend on any $\beta$.) Our first main result is that the $\beta$-dependent 
$\calg^{(n)}$ appearing in the gauge theory context \eqref{extracted} satisfy exactly
the same extended holomorphic anomaly equation. The second result is the relation between 
the orientifolded theory at $\beta=1$, and the ordinary theory at $\beta=2$ 
(see section \ref{orientifold}).

To write down the extended holomorphic anomaly equation satisfied by the 
$\calg^{(n)}$ with full $\beta$-depdendence, we need to introduce, next to the Yukawa coupling 
$C_{uuu}$, the so-called Griffiths infinitesimal invariant, 
\begin{equation}
\eqlabel{inf}
\Delta_{uu} =  \calg^{(-1)}_{uu} - C_{uu}^{\;\; \bar u}\, {\bar \calg}^{(-1)}_{\bar u}\,.
\end{equation}
This is a rank two tensor over $\calm$ whose non-holomorphicity is controlled by
\begin{equation}
\eqlabel{griffiths}
\delbar_{\bar u} \Delta_{uu} = - {C}_{uu}^{\;\; \bar u}\, {\bar \Delta}_{\bar u \bar u}\,.
\end{equation}
One can show (for example using canonical coordinates) that \eqref{griffiths} is equivalent
to the statement that $\calg^{(-1)}$ can be computed from an inhomogeneous Picard-Fuchs equation
\begin{equation}
\eqlabel{inhoPF}
\call \calg^{(-1)} = g(u)
\end{equation}
for some inhomogeneity $g(u)$, a meromorphic function over (some cover of) $\calm$. In turn,
\eqref{inhoPF} means that one can represent $\calg^{(-1)}$ as a chain integral (or open 
period) of the Seiberg-Witten differential $\lambda_{\rm SW}$ over an appropriate divisor 
on the family of curves. This is the basic geometric idea behind the extension that we will
discuss in more detail in section \ref{more}.

The extended holomorphic anomaly equation \cite{extended,tadpole}, specialized to the rigid case
\cite{real}, then reads, for $n>0$
\begin{equation}
\eqlabel{extendedhae}
\delbar_{\bar u} \calg^{(n)} = \frac 12  \sum_{\topa{n_1+n_2=n-2}{n_i\ge 0}}
{\bar C}_{\bar u}^{\; u u}\calg^{(n_1)}_u \calg^{(n_2)}_u  + 
\frac12{\bar C}_{\bar u}^{\; u u} \calg^{(n-2)}_{uu}
- {\bar \Delta}_{\bar u}^{\; u} \calg^{(n-1)}_u\,.
\end{equation}
For $n=0$, we have
\begin{equation}
\eqlabel{oneext}
\delbar_{\bar u}\del_u \calg^{(0)} = \frac 12 {\bar C}_{\bar u}^{\; u u}
C_{uuu} - {\bar \Delta}_{\bar u}^{\; u} \Delta_{uu} \,.
\end{equation}
The equations \eqref{extendedhae} and \eqref{oneext} determine the $\calg^{(n)}$ up to
certain holomorphic functions on $\calm$. To compute the complete amplitudes, one needs to 
first supply an efficient algorithm for solving the holomorphic anomaly, and then find a
sufficient number of boundary conditions at the various special points in $\calm$.

To establish our main claim, we will follow the route of solving \eqref{extendedhae} order
by order in $n$, and then showing that one may fix the holomorphic ambiguity so as to (i) 
reproduce the known results in the limit $a\to\infty$ and (ii) satisfy the expected boundary
conditions at the other special points in the moduli space.

There are various ways to solve the holomorphic anomaly equation. A convenient one is
the so-called polynomial algorithm of \cite{yayau}, described in its extended form in 
\cite{alla,komi}. (A related approach is
the ``direct integration'' of \cite{direct}.) One starts by noticing that special geometry
relates the propagator of \cite{bcov}, defined by the condition
$\delbar_{\bar u} S^{uu} = {\bar C}_{\bar u}^{\; uu}$
with the Weil-Petersson curvature on $\calm$, $\delbar_{\bar u}\Gamma_{uu}^u = 
- {\bar C}_{\bar u}^{\; uu} C_{uuu}$. So we may choose
\begin{equation}
\eqlabel{Sdef}
S^{uu} = - \frac{\Gamma^{u}_{uu}}{C_{uuu}} \,.
\end{equation}
Similarly, the terminator of \cite{extended}, characterized by
$\delbar_{\bar u} T^u = {\bar \Delta}_{\bar u}^{\; u}$
can be written as
\begin{equation}
\eqlabel{T}
T^u = - \frac{\Delta_{uu}}{C_{uuu}} \,.
\end{equation}
The main point of \cite{yayau} is then that covariant derivatives of $S^{uu}$ and $T^u$ are
known up to some holomorphic functions. In particular,
\begin{equation}
\eqlabel{consequence}
D_u S^{uu} = - C_{uuu} S^{uu} S^{uu} + f^u \,,\qquad
D_u T^u = g \,,
\end{equation}
where in fact $g(u)$ is nothing but the inhomogeneity in \eqref{inhoPF}, while $f^u(u)$ is a priori
unknown. As a consequence of \eqref{consequence}, the non-holomorphicity in the amplitudes is 
entirely through the $S^{uu}$ and $T^u$. In fact, the $\calg^{(n)}$ may be written as polynomials 
in those non-holomorphic generators with coefficients that are rational functions in $u$ (with
singularities on the discriminant locus; we count this as holomorphic).

To study the boundary conditions that fix the holomorphic ambiguity \cite{hukl1,hukl2}, one 
expands the amplitudes in the holomorphic limit \eqref{relation} and in canonical coordinates 
\eqref{general} around the special points in $\calm$, and compares with the field theory
expectations. The special loci are: the weak coupling regime, 
the monopole/dyon points (where the Yukawa coupling blows up), and the locus where $g(u)$ is 
singular.

It is convenient to encode the $\beta$-dependence of the boundary conditions via the asymptotic 
expansion of certain Schwinger integrals. These represent the contribution of integrating out 
in the general $\Omega$-background the states that are becoming light \cite{gova,nekrasov}.
We introduce two sets of functions $\Phi^{(n)}(\beta)$ and $\Psi^{(n)}(\beta)$ of $\beta$ via
\begin{equation}
\begin{split}
\int \frac{ds}{s} \frac{ \ee^{-ts}}{(\ee^{\epo s}-1)(\ee^{\ept s}-1)} &\sim
\Phi^{(0)}(\beta)\log t +
\sum_{n>0} \frac{\lambda^n}{t^n} \Phi^{(n)}(\beta) \,, \\
\int \frac{ds}{s} \frac{ \ee^{-ts}}{(\ee^{\epo s/2}-\ee^{-\epo s/2})(\ee^{\ept s/2}-\ee^{-\ept s/2})}
&\sim 
\Psi^{(0)}(\beta)\log t + 
\sum_{n>0} \frac{\lambda^n}{t^n} \Psi^{(n)}(\beta) \,.
\end{split}
\end{equation}
Here, $t\to 0$ is the mass of the state that is being integrated out, 
$\epo=\lambda\beta^{1/2}$, $\ept=-\lambda\beta^{-1/2}$, see \eqref{redef}, and we have dropped 
the most singular terms. Explicitly, for $n>0$,
\begin{equation}
\eqlabel{dumm}
\begin{split}
\Phi^{(n)}(\beta) &= (n-1)! \sum_{k=0}^{n+2} 
\frac{(-1)^k B_k B_{n+2-k}}{k! (n+2-k)!} \beta^{k-n/2-1} \,, \\
\Psi^{(n)}(\beta) &= (n-1)! \sum_{k=0}^{n+2} 
\frac{(-1)^k B_k B_{n+2-k}}{k! (n+2-k)!} 
(2^{1-k}-1)(2^{1-n-2+k}-1)  
\beta^{k-n/2-1} \,,
\end{split}
\end{equation}
where $B_k$ are the Bernoulli numbers. Notice that $B_k=0$ for $k>1$ and odd. This makes
$\Psi^{(n)}$ vanish for odd $n$. For $n=0$,
\begin{equation}
\eqlabel{respect}
\begin{split}
\Phi^{(0)} &= -\frac 14 + \frac 1{12} \beta + \frac 1{12}\beta^{-1} \,, \\
\Psi^{(0)} & = -\frac 1{24}\bigl(\beta+\beta^{-1}\bigr)  \,.
\end{split}
\end{equation}
The functions $\Phi^{(n)}$ were introduced in this context (but under different name) in 
\cite{nekrasov,neok}. The $\Psi^{(n)}$ are well-known from the study of $c=1$ string at 
radius $R=\beta$, see \eg, \cite{grkl}. The Schwinger integrals in the general $\Omega$-background 
have also been studied, for instance, in \cite{hiv,ikv,ahnt}. Notice that at $\beta=1$, we have 
for even $n$,
\begin{equation}
\Phi^{(n)}(1) =\Psi^{(n)}(1) = - \frac{B_{n+2}}{n(n+2)}  \,.
\end{equation}
Setting $n=2g-2$, we can recover the $\beta$-independent boundary conditions known from 
\cite{ghva,neok,hukl1,hukl2}. (For odd $n$, $\Phi^{(n)}(1) =\Psi^{(n)}(1)=0$).

We now briefly summarize the boundary conditions on the $\calg^{(n)}$ that we will observe
in the examples below. In the weak-coupling regime, the leading behavior of the $\calg^{(n)}$ is 
controlled by the perturbative contribution to the partition function \eqref{central}, and 
contains the functions $\Phi^{(n)}(\beta)$ \cite{nekrasov,neok} (see appendix). At 
monopole/dyon points, we will find,
extending the gap structure of \cite{hukl1}, that $\calg^{(n)}$ have a leading singularity
$\sim t^{-n}$, followed by regular terms $\calo(t^0)$. Quite interestingly, we will find that 
the leading ($\beta$-dependent) coefficient is sometimes governed by the $\Psi^{(n)}$, sometimes
by the $\Phi^{(n)}$, and sometimes by an as yet unidentified function. Finally, we find that the 
$\calg^{(n)}$ are regular at points where $g(u)$ is singular, but which are not monopole/dyon 
points.

\section{Examples}

In this section, we study in detail the partition function $Z(a,\epo,\ept;q)$ for
$SU(2)$ gauge theory coupled to $N_f\le 3$ massless fundamental hypermultiplets. These are
the simplest models with a one-dimensional Coulomb branch and an extensive literature. 
We have summarized in the appendix the results from instanton counting \cite{nekrasov}. 
In the B-model, it suffices for the moment to recall the
Picard-Fuchs differential operators written in terms of the global coordinate $u$.
The special geometry of the moduli space has a discrete symmetry $\zet_{1/\alpha}$ that allows 
us to write the differential equation in terms of the coordinate $z= u^{1/\alpha}$.
Here, $\alpha$ is given in the following table
\begin{equation}
\begin{array}{c|c|c|c|c}
N_f & 0 & 1 & 2 & 3 \\\hline
\alpha& \frac 12 & \frac 13 & \frac 12 & 1  
\end{array} \,.
\end{equation}
Moreover, for massless hypermultiplets, the constant solution of the third order Picard-Fuchs 
equation decouples from the monodromies, so that we may work with the simpler second order 
operator given by
\footnote{We work with a strong-coupling scale such that the discriminant contains the
locus $u^{1/\alpha}=1$. On the A-side, we work at $q=1$.}
\begin{equation}
\eqlabel{genPF}
\call_\alpha = \theta(\theta - \alpha) - z \bigl(\theta-\frac{\alpha}{2}\bigr)^2 \,,
\end{equation}
where $\theta= z \frac{d}{dz}$. We also record that the general solution of the differential
equation $\call_\alpha\varpi(z)=0$ around $z=\infty$ can be obtained from
\begin{equation}
\eqlabel{encode}
\varpi(z;H) = \sum_{n=0}^\infty \frac{\Gamma(n+H+\alpha/2)\Gamma(n+H-\alpha/2)}
{\Gamma(n+H+1)^2} z^{\alpha/2-n-H}
\end{equation}
as
\begin{equation}
\varpi_0 = \varpi(z;H=0)\sim z^{\alpha/2} + \cdots \,,\qquad
\varpi_1 = \del_H \varpi(z;H=0) \sim -\varpi_0 \log z + \cdots \,.
\end{equation}
The flat coordinate at $u\to\infty$ is $a\propto \varpi_0\sim u^{1/2}$, and matching 
the asymptotic behaviour of the periods with the perturbative computation in the gauge
theory determines the proper linear combinations yielding $a_D= \del_a\calf^{(0)}$ and 
the prepotential. The precise coefficients are not important for our purposes, as only 
the Yukawa coupling enters the recursion relations.

A common feature of the three models is the singularity at $1-z=1-u^{1/\alpha}=0$.
In the coordinate $\tilde z = 1-z$, with $\tilde\theta = \tilde z\frac{d}{d\tilde z}$
the Picard-Fuchs operator takes the form 
\begin{equation}
\call_\alpha = \bigl(\tilde z^{-1}-1\bigr) \tilde\call_\alpha \,,
\end{equation}
with
\begin{equation} 
\tilde\call_\alpha = \tilde \theta (\tilde\theta-1) - \tilde z \bigl(\tilde\theta -
\frac{\alpha}2\bigr)^2 \,.
\end{equation}
The solutions near $\tilde z=0$ can be encoded similarly as in \eqref{encode}. In most
other respects, each of the four models we consider is special, so we now have to split 
the discussion. In each case, we start with a look at the expansion of the instanton
partition funtion to learn whether we should use the standard or extended holomorphic 
anomaly equation, whether discrete symmetries are broken or not, \etc. This also dictates 
the ansatz we make for the holomorphic ambiguity. This information, together with
the perturbative and one-instanton contribution is summarized in the 
appendix (see especially, \eqref{perturbative} and \eqref{oneinst0}, \eqref{oneinst1}, 
\eqref{oneinst2}, \eqref{oneinst3}).

\paragraph{\texorpdfstring{Pure gauge theory}{Pure gauge theory}}

The Yukawa coupling of the model is given by
\begin{equation}
C_{uuu} = \frac{2}{1-u^2} \,,
\end{equation}
and for approriate normalization of the periods $a$ and $a_D$, we obtain
\begin{equation}
\del^2_a a_D = C_{aaa} = \Bigl( \frac{\del u}{\del a} \Bigr)^3 C_{uuu} =
-\frac{8}{a}-\frac{12}{a^5}-\frac{105}{4 a^9}-\frac{495}{8 a^{13}}
+\cdots \,,
\end{equation}
thus matching the known results, which are those of the instanton counting after setting 
$q=1$ (see appendix). We also note that in our integration scheme, we have quite simply 
$f^u= -1/8$.

To evaluate the $\beta$-dependent higher order terms, we may use the ordinary holomorphic 
anomaly equation \eqref{hae}, since terms of higher odd order in the expansion \eqref{extracted}
vanish identically (see appendix). But to be more systematic, we continue to use the 
parameterization via the $\calg^{(n)}$. At one loop, we find \cite{mowe,nekrasov,nayo1}
\begin{equation}
\calg^{(0)}(\beta) = \frac 12 \log {\rm Im}\tau - \frac{\beta+\beta^{-1}}{24}\log |1-u^2|^2 \,,
\end{equation}
also reproducing the instanton counting results. In higher order (even $n$), we solve 
\eqref{hae} using the direct integration algorithm, to find $\calg^{(n)}$ as a polynomial 
in the non-holomorphic propagator $S^{uu}$ with coefficients rational functions of $u$, 
up to the constant term $A^{(g)}(u;\beta)$. Constraints on the asymptotic behaviour 
at $u\to\infty$ and $u=\pm 1$ require the ansatz
\begin{equation}
\eqlabel{impose}
A^{(g)}(u;\beta) = \frac{u^{3n/2}}{(1-u^2)^{n}} \sum_{i=1}^{n-1} P^{(n)}_i(\beta) u^{-2i} \,.
\end{equation}
We may fix the coefficient functions, $P^{(n)}_i(\beta)$, by imposing the gap structure at 
the monopole/dyon points $z=1$. The local coordinate there is $a_D\sim \tilde z$,
and we have
\begin{equation}
\eqlabel{completely}
\calg^{(n)} = \frac{\Psi^{(n)}(\beta)}{a_D^n} + \calo(1) \,,
\end{equation}
with $\Psi^{(n)}$ given in \eqref{dumm}. Counting parameters, there are at each order $n$ 
unknown functions $P^{(n)}_i(\beta)$, which are precisely determined by as many conditions 
from \eqref{completely}.

This way of fixing the holomorphic ambiguity is the generalization of \cite{hukl1} to general
values of $\beta$. One may check that it is in agreement with the expectations. In particular, 
the leading order term in the weak-coupling expansion takes the form
\begin{equation}
\calg^{(n)} = \frac{ \Phi^{(n)}(\beta)}{2^{n-1} a^n} + \cdots  \,,
\end{equation}
and the one-instanton sector is matched as well. Moreover, one can verify that the amplitudes
are regular around $z=0$.

\paragraph{\texorpdfstring{$N_f=1$}{Nf=1}}

The theory with one flavor is the most interesting one from the present point of view, 
as here the extended
holomorphic anomaly equation gears up to its full power. But first we note the Yukawa
coupling
\begin{equation}
C_{uuu} \propto \frac{u}{1-u^3} \,,
\end{equation}
which matches the term of order $\lambda^{-2}$ from instanton counting. The term at order
$\lambda^{-1}$ is non-zero in this model. For $q=1$, we have
\begin{equation}
\eqlabel{see}
\calt= \frac{\calg^{(-1)}}{\beta^{1/2}-\beta^{-1/2}} = 
\frac a4 -\frac{1}{4 a^2}+\frac{7}{384a^8}-\frac{1131}{163840 a^{14}}+
\frac{3705}{917504a^{20}}+\cdots\,.
\end{equation}
Plugging in $a=a(z)$, and acting with the Picard-Fuchs operator, we find
\begin{equation}
\eqlabel{plugging}
\call_{1/3}\calt(z) = \frac 1 3z^{2/3} \,.
\end{equation}
We defer further discussion of the geometric origin of this inhomogeneity to section \ref{more},
and proceed with the integration of the extended holomorphic anomaly equation. At one-loop,
we have
\begin{equation}
\del_u \calg^{(0)} = \frac 12 S^{uu} C_{uuu} + \frac 12 C_{uuu} (T^u)^2 +
\frac{\beta+\beta^{-1}}{24}\frac{3 u^2}{1-u^3}
+ \frac{\bigl(\beta^{1/2}-\beta^{-1/2}\bigr)^2}{8}\frac 1u \,.
\end{equation}
At higher order, the holomorphic ambiguity is parameterized by
\begin{equation}
\eqlabel{full}
A^{(n)}(u;\beta) =
\begin{cases}
\displaystyle \frac{u^{5n/2}}{(1-u^3)^n}\sum_{i=0}^{n-1} P_i^{(n)} u^{-3i} 
+ \frac 1{u^{2n}}\sum_{j=0}^{n/2} Q_j^{(n)} u^{3j} & \text{$n$ even} \\
\displaystyle
\frac{u^{(5n-9)/2}}{(1-u^3)^{n-1}}\sum_{i=0}^{n-2} P_i^{(n)} u^{-3i}
+ \frac{1}{u^{2n}}\sum_{j=0}^{\frac{n-1}2} Q_j^{(n)} u^{3j} & \text{$n$ odd}
\end{cases}  \;\;.
\end{equation}
This ansatz is slightly redundant, but more intuitive than the minimal one: The
prefactor $(1-u^3)^{-n}$ captures the leading singularity at the monopole point,
while the prefactor $u^{-2n}$ is explained from the leading behaviour of the
solution around $u=0$. Indeed, inspecting the Picard-Fuchs equation, the flat 
coordinate there behaves as $a_0\sim u$, and the extension as $\calt\sim u^2\sim a_0^2$.
The maximal order of a singularity is $\calt^{-n}$.
The summation ranges are dictated by the condition that the ambiguities not spoil
each other's asymptotic behaviour. In particular, around $u\to\infty$, we should
have the behaviour $\sim a^{-n}\sim u^{-n/2}$ for $n$ even, and $\sim a^{-n-3}
\sim u^{-(n+3)/2}$ for $n$ odd.

To fix the coefficient functions $P_i(\beta)^{(n)}$, $Q_j^{(n)}(\beta)$ in \eqref{full},
we have $n$ conditions from the gap structure at the monopole points $u^3=1$
(the $\zet_3$ symmetry remains unbroken)
\begin{equation}
\calg^{(n)}= \frac{\Psi^{(n)}(\beta)}{a_D^{n}} + \calo(1) \,.
\end{equation}
(Note that this is regular for $n$ odd.) We also find that the $\calg^{(n)}$ are
regular at $u=0$, accounting for $\sim 3n/2$ conditions. Regularity at $u=0$ despite
the vanishing of $\calt$ (singularity in $\Delta_{uu}$) is reassuring, since we 
would not know how to explain any 
singularity from integrating out a massless state. This is in contrast to the application 
of the extended holomorphic anomaly in the context with background D-branes 
\cite{extended,tadpole,real}, where the vanishing of $\calt$ signals a tensionless
domain wall, and leads to so far uncontrolled singularities in the higher loop amplitudes.

After fixing the holomorphic ambiguities in this way, we can check the expansion around
weak coupling with the results from instanton counting, finding complete agreement.

\paragraph{\texorpdfstring{$N_f=2$}{Nf=2}}

At first sight, the two-flavored case appears somewhat uninteresting, since the special
geometry is so closely related to that of the pure gauge theory. In particular, 
$\alpha=1/2$ in both cases, and the Yukawa coupling,
\begin{equation}
C_{uuu} = \frac 1{4(1-u^2)} \,,
\end{equation}
and $f^u=-1$ differ only in the normalization of $u$.

The surprise, however, appears when we look at the loop amplitudes. Already for $n=0$,
we find that turning on the $\beta$-deformation away from $\beta=1$, {\it breaks the
$\zet_2$ symmetry} between the monopole and dyon point at $u=+1$ and $u=-1$, respectively.
Indeed, we find
\begin{equation}
\del_u \calg^{(0)} = \frac 12 S^{uu} C_{uuu}  
+ \frac{\beta- 3 + \beta^{-1}}{6} \frac{1}{1+u}
+ \frac{\beta+\beta^{-1}}{12}\frac{1}{1-u} \,.
\end{equation}
We recognize in these expressions the leading behaviour at the two components of the
discriminant locus corresponding to $u=-1$ and $u=+1$ to be $2\Phi^{(0)}(\beta)$ and 
$2\Psi^{(0)}(\beta)$ from eq.\ \eqref{respect}. At $\beta=1$, the $\zet_2$ symmetry is 
restored, and we recover the known results \cite{hukl2}.

This structure persists at higher order as well. Amplitudes at odd $n$ are zero. For
even $n$, we work with the ansatz
\begin{equation}
A^{(n)}(u;\beta) = \frac{u^{n/2}}{(1+u)^n} \sum_{i=0}^{n-1} P^{(n)}_i u^{-i} +
\frac{u^{n/2}}{(1-u)^n} \sum_{j=0}^{n-1} Q^{(n)}_j u^{-j} 
\end{equation}
for the holomorphic ambiguity.
Then the gaps at $u=\pm1$,
\begin{equation}
\calg^{(n)} =
\begin{cases}
\displaystyle
\frac{2\Phi^{(n)}}{a_{D,+}^n} +\calo(1) & \text{around $u=+1$} \\
\displaystyle
\frac{2\Psi^{(n)}}{a_{D,-}^n} +\calo(1) & \text{around $u=-1$}
\end{cases} 
\end{equation}
are sufficient to completely fix the $P^{(n)}_i$, $Q^{(n)}_j$, also in this case.

\paragraph{\texorpdfstring{$N_f=3$}{Nf=3}}

Our last example is an interesting mix of all the previous ones. The $u$-plane has
no discrete symmetry, and the second component of the discriminant locus moves to $u=0$. 
We have
\begin{equation}
C_{uuu} = \frac{1}{64u(1-u)}
\end{equation}
and $f^u=-16$. As for $N_f=1$, the amplitudes at odd order are generally non-zero. The 
term at $n=-1$, however, is just $\calg^{(-1)} = (\beta-\beta^{-1/2})(-\frac a4+\frac 14)$ 
and as a consequence, $\Delta_{uu}=0$. This simplifies the integration scheme, 
but we must still use the extended holomorphic anomaly equation. The one-loop amplitude 
comes out to be
\begin{equation}
\del_u\calg^{(0)} = \frac 12 S^{uu} C_{uuu} + \frac{\beta+\beta^{-1}}{24}\frac{1}{1-u}
+ \frac{5\beta-18+5\beta^{-1}}{24}\frac{1}{u} \,.
\end{equation}
The term at $n=1$ is purely holomorphic, and we find
\begin{equation}
\calg^{(1)} = \bigl(\beta^{1/2}-\beta^{-1/2}\bigr)^3 \frac 1 u \,.
\end{equation}
More generally, we have a holomorphic ambiguity
\begin{equation}
A^{(n)}(u;\beta) = 
\begin{cases}
\displaystyle
\frac{u^{n/2}}{(1-u)^n} \sum_{i=0}^{n-1} P^{(n)}_i u^{-i} + 
\frac{1}{u^n} \sum_{j=0}^{n/2} Q^{(n)}_j u^j&\text{$n$ even} \\
\displaystyle
\frac{u^{(n-1)/2}}{(1-u)^{n-1}}\sum_{i=0}^{n-2} P^{(n)}_i u^{-i}+
\frac{1}{u^n} \sum_{j=0}^{(n-1)/2} Q^{(n)}_j u^j & \text{$n$ odd}
\end{cases} \;\;.
\end{equation}
As for the boundary conditions, we find at $u=+1$ a familiar gap structure
\begin{equation}
\calg^{(n)}= \frac{\Psi^{(n)}}{a_D^n} + \calo(1) \,.
\end{equation}
In distinction to $N_f=1$, $u=0$ is not a regular point (already at $\beta=1$). 
We also find a gap, 
\begin{equation}
\calg^{(n)} = \frac{X^{(n)}}{a_0^n} + \calo(1)\,,
\end{equation}
with leading coefficients $X^{(n)}(\beta)$ that curiously are non-zero also for 
odd $n$ (but vanish there at $\beta=1$). The first few are
\begin{equation}
\begin{split}
X^{(0)} &= \frac{5\beta-18+5\beta^{-1}}{24}\,, \quad
X^{(2)} = \frac{-67\beta^2+540 \beta -1330 +540 \beta^{-1} -67 \beta^{-2}}{5760}\,,
\\
X^{(1)}  &= \frac{(\beta^{1/2}-\beta^{-1/2})^3}{8}\,,\quad
X^{(3)}  = \frac{(\beta^{1/2}-\beta^{-1/2})^3 (\beta+6+\beta^{-1})}{96} \,.
\end{split}
\end{equation}

\section{More on the Extension}
\label{more}

Our development of the B-model formalism for general $\beta$ in secion \ref{develop} 
involved in a central fashion the ``extension'' of special geometry by the term at order
$\lambda^{-1}$ in the expansion of $\log Z$,
\begin{equation}
\calg^{(-1)} = \bigl(\beta^{1/2}-\beta^{-1/2}\bigr) \calt \,.
\end{equation}
In this section, we show how $\calt$ can be recovered from the Seiberg-Witten
geometry.

As a motivation, we recall that in the context of topological strings with D-branes
and orientifolds \cite{extended,tadpole}, $\calt$ is the topological disk (or disk+crosscap) 
amplitude.
It can be written in terms of the holomorphic Chern-Simons functional, or as an
integral $\sim\int^C \Omega$ of the holomorphic three-form of the Calabi-Yau over a 
three-chain ending on a holomorphic curve $C$ that represents the background D-brane.
See ref.\ \cite{normal} for the relevant Hodge theoretic notions. The reduction
of the holomorphic three-form to the present context is the Seiberg-Witten differential 
$\lambda_{\rm SW}$ on the curve $\calc_u$, and the holomorphic curve $C$ becomes a 
pair of points $p_-$, $p_+$, varying holomorphically with $\calc_u$ as a function of $u$. 
Hence we expect that for appropriate choice of $p_{\pm}$, we have the representation
\begin{equation}
\eqlabel{represent}
\calt = \int_{p_-}^{p_+} \lambda_{\rm SW} \,.
\end{equation}
An important caveat is in order. In distinction to the holomorphic three-form of a 
Calabi-Yau threefold, the Seiberg-Witten differential is not unique. It is merely
characterized by the condition that $\del_u\lambda_{\rm SW}=\omega_u$ be the 
holomorphic one-form of the elliptic curve, {\it up to exact terms}. Modifying
$\lambda_{\rm SW}$ by an exact form will change integrals such as \eqref{represent}. 
(Similar ambiguities play a role in recent studies of surface operators in $\caln=2$ 
gauge theory, see, \eg, \cite{aggtv,kpw,dgh}.) The invariant Seiberg-Witten geometry
capturing the refinement should thus involves the curve, the differential, and
the points $p_{\pm}$. 

From our examples, the only case where we can ask the question in an invariant
way is $N_f=1$. The Seiberg-Witten curve may be written as \cite{sw2},
\begin{equation}
\calc_u:\;\; y^2 = x^2(x-u) + \frac{4}{27}
\end{equation}
(we chose $\Lambda$ such that the discriminant is at $u^3=1$),
and the differential as
\begin{equation}
\lambda_{\rm SW} = \sqrt{3}\frac{dy}{x} = \sqrt{3}\frac{2u-3x}{y} dx \,.
\end{equation}
We claim that with respect to these choices, the correct combination of points can
be obtained by intersecting the curve with the plane $x-u=0$, namely 
\begin{equation}
\eqlabel{special}
p_{\pm}: (x,y) = \bigl(u,\pm \sqrt{\frac{4}{27}}\bigr) \,.
\end{equation}
There are various ways to check this claim. The most straightforward is to directly
integrate. Indeed, the expansion around $u\to\infty$,
\begin{equation}
\int_{p_-}^{p_+} \lambda_{\rm SW} = \frac{4}{3u} + \frac{32}{243 u^4} + 
\frac{512}{10935 u^7} + \cdots
\end{equation}
matches (up to a rational period) the solution of the inhomogeneous Picard-Fuchs 
equation \eqref{plugging}.

Note that this discussion did not explain whether the choice of points \eqref{special} 
(with respect to the choice of differential) was distinguished in any sense. The formalism 
of section
\ref{develop} would go through for any other reasonable choice as well, but only
\eqref{special} leads to the correct answer. Lacking a deeper understanding, we can
point out one way in which one might attempt to understand the invariant physical meaning 
of $\calt$.\footnote{The following observations arose in a conversation with Samson 
Shatashvili.}

In the work \cite{nesh}, the four-dimensional gauge theory in the $\Omega$-background
with $\ept=0$ was studied, and it was shown that this theory provides the quantization of
the classical integrable system underlying the original four-dimensional theory. In 
particular, it was shown that the twisted superpotential
$\calw(a,\epo;q) = \lim_{\ept\to 0}  \ept \log Z(a,\epo,\ept;q)
$
can be identified with the Yang-Yang function of the integrable system. Moreover, 
the remaining parameter $\epo$ is identified with
the Planck constant of the quantization procedure. We may relate the expansion in $\epo$
to our parametrization \eqref{extracted} via
\begin{equation}
\calw(a,\epo;q) = \frac{1}{\epo} \calf^{(0)} + \calt + \calo(\epo)
\end{equation}
Viewed from this angle, $\calt$ is nothing but a one-loop term, which shows its 
special role. Moreover, one should be able to compute it directly in the semi-classical 
expansion of the relevant integrable system. We leave this line of investigation for the 
future.

\section{Orientifold}
\label{orientifold}

In this section, we explain a relation between two special sets of values for the
parameters of the $\Omega$-background. The essential message is that, at least in 
certain cases, the theory at $\epo=-2\ept$ can be viewed, in a very precise sense,
as the orientifold of the theory at $\epo=-\ept$. In a way, this result has been
anticipated by the relation between the Nekrasov deformation of gauge theories and
the $\beta$-ensemble of generalized matrix models \cite{diva}. Indeed, it is well-known
that when the value $\beta=-\epo/\ept=1$ corresponds to the $U(N)$ matrix models,
then $\beta=1/2,2$ correspond to $SO(N)$, $Sp(N)$, respectively, which are just the 
orientifolds of $U(N)$. (The change of the string coupling $\lambda=-\epo\ept \sim 1/N$, 
is also accounted for in this relationship).

We uncover the relation between $\beta=1$ and $\beta=2$ purely from the instanton 
counting in the gauge theory. This has several virtues. First of all, we see the
remarkable cancellation, as in the orientifold we sum only over a very small subset 
of the ``real'' instantons, but still recover the result of the full sum for 
different value of the parameters. Second, the identification of the 
$\beta$-parameter with the radius $R=\beta$ of the $c=1$ string, and the comparison
with the moduli space of $c=1$ conformal field theories \cite{dvv}, is suggestive
of the possible existence of an entire new branch of topological theories: The value 
$\beta=2$ is precisely the point where the orbifold branch of $c=1$ theories 
touches the circle branch. It would be interesting to see how to move in this
direction. Finally, we will see that the relationship does not strictly hold in all 
cases. Namely, it fails us for an odd number of flavors. We attribute this to the
accidental (in)completeness of the orientifold prescription as the right quotient. 
We also explain some parallels with orientifolds of the topological string.

\paragraph{The squareroot}

The basic idea is quite simple, and indeed nothing else than a new manifestation of
the ``real topological string principle'', developped in 
\cite{bouchard1,bouchard2,tadpole,real,kp}. In the localization computation of
\cite{nekrasov}, the instanton counting partition function is written as the sum 
of contributions from the fixed points of the certain group action on the moduli 
space of instantons. We show only the parameters $\epo,\ept$ related to $\T^2$ 
action on $\reals^4 \cong \complex^2\ni (z_1,z_2)\to (\ee^{\ii \epo} z_1,
\ee^{\ii \ept}z_2)$.
\begin{equation}
Z^{\it inst}(\epo,\ept) = \sum_{Y} R_Y(\epo,\ept) \,,
\end{equation}
where $Y$ some colored partitions label the fixed points, and $R_Y(\epo,\ept)$ is 
a rational function of the various parameters (see appendix for the examples). 

Consider now the real structure on $\complex^2$: $\sigma: (z_1,z_2)\to 
(\bar z_2,\bar z_1)$. This commutes with a one-dimensional subtorus of $\T^2$, so 
localization still applies. In fact the subtorus is nothing else than $\epo=-\ept$, 
\ie, the anti-diagonal $U(1)$. 
We also assume an appropriate lift to the rest of the data. In particular, $\sigma$ 
acts on the set of $Y$'s. The real topological string principle then instructs us to 
consider the sum over the invariant configurations,
\begin{equation}
\eqlabel{signs}
Z^{\it real\; inst}(\epo) = \sum_{Y=\sigma(Y)} \sqrt{R_Y(\epo,-\epo)} \,,
\end{equation}
exploiting the fact that $R_Y(\epo,-\epo)$ is a perfect square.
We have implemented this procedure for $SU(2)$ gauge theory with $N_f\le 4$ flavors. 
(Of course, \eqref{signs} requires the specification of a sign for each invariant 
fixed point. Luckily, this is quite straightforward in the present case.) 
Our main result is not so much that the resulting expression makes sense (for 
example, in having a sensible limit as $\epo\to 0$; this is true also for 
$N_f=1,3$). Rather, we find that for $N_f=0,2,4$,
\begin{equation}
\eqlabel{exciting}
Z^{\it real\; inst}(\epo) = Z^{\it inst}(\epo,-2\epo) \,,
\end{equation}
which can be checked in the appropriate expansion.

\paragraph{Return of the topological string}

It is well-known that the $\caln=2$ gauge theory can be embedded into string 
theory via geometric engineering \cite{klmvw,kkv}. The gravitational corrections,
$\calf^{(g)}=\calg^{(g)}(\beta=1)$, computed in the $\Omega$-background at 
$\ept=-\epo$ are then identified with the field theory limit of the genus-$g$
topological string amplitudes \cite{nekrasov}. It is an important
question whether the main results of the present paper, namely, extension of the 
holomorphic anomaly, and embedding of orientifold at $\beta=2$, can be lifted 
to the full-fledged topological string on any Calabi-Yau threefold.

At the moment, we can offer one further piece of evidence that at least some
aspect will survive. The Calabi-Yau's of geometric engineering are toric, and
we can compute the partition functions in the topological vertex formalism
\cite{akmv}, and also their orientifolds \cite{kp}. \footnote{It can, for 
instance, be checked that the real topological string on 
$\projective^1\times\projective^1$ computed from the real vertex coincides
with the termwise squareroot of Nekrasov's five-dimensional $SU(2)$ partition 
function (similarly, for the geometries that engineer $N_f=2,4$).}
In ref.\ \cite{real}, the real vertex was applied to the local $\projective^2$ 
geometry (which incidentally is not even an engineering geometry), and the results 
were compared with those of the holomorphic anomaly 
equation. In particular, expanding around the conifold point of local $\projective^2$, 
it was found that the genus-$g$ Klein bottle amplitudes
\footnote{The $\calg^{(2g-2)}$ featuring here don't depend on any $\beta$.} 
\begin{equation}
\eqlabel{appropriate}
\calk^{(g)} = \calg^{(2g-2)} - \calf^{(g)}
= \frac{\psi^{(g)}}{a_D^{2g-2}} + \calo(1)
\end{equation}
show the universal gap structure.
The first few coefficients were found to be (up to the model-dependent factor of 
$3^{g-1}$)
\begin{equation}
\eqlabel{found}
\psi^{(2)}=-\frac{3}{128},\,\,\, \psi^{(3)}=\frac{9}{512},\,\,\, \psi^{(4)}=-\frac{157}{4096} \,,
\end{equation}
but not otherwise identified in \cite{real}. Using these results, we can test the present 
idea that the topological orientifolds can be embedded in a putative refinement of the
theory at $\beta=2$ by comparing with the universal behaviour found in the gauge theory 
examples. Taking the appropriate linear combination \eqref{appropriate}, we predict%
\footnote{The relative factor of $2^g$ is due to the redefinition of the
string coupling in the orientifold.}
\begin{equation}
\psi^{(g)} = 2^{g}\Psi^{(2g-2)}(2) -  \Psi^{(2g-2)}(1) = 
\frac{1}{2^{2g+1} g(g-1)} ((2^{2g}-1)B_{2g}-g E_{2g-2})  \,,
\end{equation}
where $E_g$ are Euler numbers. This indeed matches the coefficients \eqref{found} found
in \cite{real}.

\section{Conclusions}

In this paper, we have uncovered some new properties of the partition function of
$\caln=2$ gauge theory in the $\Omega$-background, and found various hints that
those structures can be lifted to the topological string. Many interesting 
physical and mathematical questions remain, which we hope to take up in the
near future.

\begin{acknowledgments}
We thank Can Koz\c caz, Wolfgang Lerche, Sara Pasquetti, and especially Samson Shatashvili
for valuable dicussions and comments. J.W.\ thanks Sergei Gukov, Hiraku Nakajima, 
and Nikita Nekrasov for some early discussions (2007/8) on the main problem addressed here. 
D.K. likes to thank CERN-TH for hospitality where this work has been initiated. The work of 
D.K. was supported by the WPI initiative by MEXT of Japan. 
\end{acknowledgments}
  
\appendix

\section{Nekrasov's formulae}
\label{Neksec}

In this appendix, we will briefly recall the instanton calculation of \cite{nekrasov} 
and we collect some key observations regarding the structure of the resulting partition 
function for $SU(2)$ gauge group with up to four flavors.

\paragraph{Basics}

Consider the (compactified) moduli space $\Mcal_k$ of $k$-instantons of $U(N)$ gauge theory 
with $N_f$ fundamentals in $\R^4\cong \complex^2$. According to \cite{lns1,lns2,nekrasov}, 
the corresponding 
instanton partition function, denoted as $Z^{inst}$, can be calculated via localization with 
respect to the $U(N)\times U(N_f)\times\T^2$ group action on $\Mcal_k$, where $\T^2$ is the 
maximal torus of the $SO(4)$ rotation group of $\R^4$. Namely,
\beq\eqlabel{NekZinstL}
Z^{\it inst}(\vec a,\vec \m,\epsilon_1,\epsilon_2;q)=\sum_k q^k 
\int_{\Mcal_k}{\bf e}(V\otimes\C^{N_f})\,,
\eq
where $q$ is a parameter, $\vec a=(a_1,a_2,\dots,a_N)$ are coordinates of the Cartan subalgebra 
of $U(N)$, $\vec \m=(\m_1,\m_2,\dots,\m_{N_f})$ the masses of the $N_f$ fundamentals (coordinates 
on the Cartan subalgebra of the flavor group $U(N_f)$), $\epsilon_i$ are the coordinates on 
the Lie algebra of $\T^2$, ${\bf e}$ denotes the equivariant Euler class with respect to the 
group $U(N)\times U(N_f)\times\T^2$ and $V$ is the bundle over $\Mcal_k$ of solutions of the 
Dirac equation in the instanton background.

The partition function \req{NekZinstL} can be expressed as follows, which is convenient for 
explicit computation \cite{nekrasov}
\beq\eqlabel{NekZinst}
Z^{\it inst}(\vec a,\vec \m,\epsilon_1,\epsilon_2;q)=\sum_{\vec Y}\frac{\prod_{i=1}^{N_f}
\prod_\gamma f_\gamma^{\vec Y}(\m_i)}{\prod_{\alpha,\beta}n_{\alpha,\beta}^{\vec Y}}
q^{|\vec Y|}\,,
\eq
where $\vec Y=(Y_1,Y_2,\dots,Y_N)$ is an $N$-tuple of partitions (Young diagrams), $q$ a 
parameter, $n_{\alpha,\beta}^{\vec Y}$ collects the contribution from the gauge sector 
(we use the form presented in \cite{nayo1})
\beq
n_{\alpha,\beta}^{\vec Y}=\prod_{s\in Y_\alpha}(-l_{Y_\beta}(s)\epsilon_1 +
(a_{Y_\alpha}(s)+1)\epsilon_2+a_{\beta\alpha})\prod_{t\in Y_\beta}((l_{Y_\alpha}(t)+1)
\epsilon_1 -a_{Y_\beta}(t)\epsilon_2+a_{\beta\alpha})\,,
\eq
with $s$ and $t$ running over all boxes $(i,j)$ in $Y_\alpha$ and $Y_\beta$, respectively,
$a_{Y}(i,j):=\mu_i^{Y}-j$, $l_Y(i,j):={\mu_j^{Y^t}}-i$ ($\mu_i^Y$ and ${\mu_i^{Y^t}}$ 
denote the $i$-th column of the Young diagram $Y$ and its transpose, respectively), 
$f_\gamma^{\vec Y}(\m)$ represents the contribution from a single fundamental of mass $\m$, 
\ie,
\beq
f_\gamma^{\vec Y}(\m)=\prod_{i=1}^{\mu_1^{Y^t_\gamma}}\prod_{j=1}^{\mu_i^{Y_\gamma}}
\left( a_\gamma+\m+ \epsilon_1(i-1)+\epsilon_2(j-1)\right) \,,
\eq
$a_{\beta_\alpha}=a_\beta-a_\alpha$, and indices $\alpha,\beta,\gamma$ running over 
$1,\dots,N$. Note that one can restrict to $SU(N)\subset U(N)$ by enforcing $\sum_i a_i=0$.

The instanton partition function \req{NekZinstL} must be supplemented by a perturbative part, 
denoted as $Z^{\it pert}$, to yield the full partition function in the $\Omega$-background,
\beq\eqlabel{NekZ}
Z(\vec a,\vec \m,\epsilon_1,\epsilon_2;q)=Z^{\it pert}(\vec a,\vec \m,\epsilon_1,\epsilon_2)\,
Z^{\it inst}(\vec a,\vec \m,\epsilon_1,\epsilon_2;q)\,.
\eq
We also give some details on the perturbative part of \req{NekZ}. Namely, one takes 
\cite{nekrasov,neok,nayo2},
\beq
\log Z^{\it pert}(\vec a,\vec \m,\epsilon_1,\epsilon_2)=\sum_{\alpha,\beta} 
\gamma_{\epo,\ept}(a_{\alpha\beta})-\sum_{\gamma}\sum_{i=1}^{N_f}  
\gamma_{\epo,\ept}(a_\gamma+\m_i)\,,
\eq
with
\beq
\eqlabel{gammaep}
\gamma_{\epo,\ept}(x)=\sum_{n=3}^\infty \frac{c_n(\epo,\ept)}{n(n-1)(n-2)} x^{2-n}+\mathcal O(1)\,,
\eq
and $c_n$ defined via the expansion
\beq\eqlabel{cdef}
\frac{1}{(e^{\epo t}-1)(e^{\ept t}-1)}=\sum_{g=0}^\infty \frac{c_g(\epo,\ept)}{g!} t^{g-2}\,.
\eq
The lower order terms in \req{gammaep} are not so important for our considerations and therefore 
we omit to display them explicitly. Especially, in the $SU(2)$ case with $N_f$ massless flavors 
(we drop the $\vec m$ parameter), we infer
\begin{equation}
\eqlabel{perturbative}
\log Z^{\it pert} \sim \sum_n \lambda^n \frac{(1-2^{n}N_f)}{2^{n-1}a^n}\Phi^{(n)}(\beta)  \,,
\end{equation}
where as in the main text, $\beta= -\epo/\ept$, and $\Phi^{(n)}(\beta)$ are functions
defined in \eqref{dumm}.

\paragraph{Observations}

When expanding $Z$ as in eq.\ \eqref{extracted}, we have to remember to first expand
$\log Z$ in powers of $q$. We may then set $q=1$. We consider $SU(2)$ gauge theory with
$N_f\le 4$.

\subparagraph{Pure $SU(2)$:}

We set $a_1=-a_2$ and drop the empty $\vec \m$ parameter. Terms at odd $n$
vanish, except for $n=-1$, where we have $\calg^{(-1)}=(\beta^{1/2}-\beta^{-1/2})\frac a2$,
see \cite{nayo2}. For the one-instanton sector, we obtain
\beq
\eqlabel{oneinst0}
\log Z^{\it inst}(a,\beta^{1/2}\lambda,-\beta^{-1/2}\lambda)|_{q^1}=-\sum_{i=0}^\infty 
\frac{(\beta-1)^{2i}}{2^{2i+1}\beta^{i} } \frac{\lambda^{2 i-2}}{a^{2i+2}}=:
\Fcal^{\rm even}_1\,.
\eq

\subparagraph{$N_f=1$:}
 
For a single flavor of mass $\m$, terms at odd powers of $\lambda$ are non-zero,
and already $\calg^{(-1)}$ is highly non-trivial. See \eqref{see} for the explicit
expression when $\m=0$. The one-instanton sector is
\beq
\eqlabel{oneinst1}
\log Z^{\it inst}(a,\m,\beta^{1/2}\lambda,-\beta^{-1/2}\lambda)|_{q^1}= 
\sum_{i=0}^\infty 
\frac{(\beta-1)^{2i+1}}{2^{2i+2}\beta^{i+1/2}}\frac{\lambda^{2i-1}}{a^{2i+2}} +
\m\,\Fcal^{\rm even}_1 =: \Fcal^{\rm odd}_1 + \m \Fcal^{\rm even}_1\,.
\eq

\subparagraph{$N_f=2$:}

We have two flavors with masses $\vec \m=(\m_1,\m_2)$. The $1$-instanton sector reads
\beq
\eqlabel{oneinst2}
\log Z^{\it inst}(a,\vec \m,\beta^{1/2}\lambda,-\beta^{-1/2}\lambda)|_{q^1}=(\m_1+\m_2)\, 
\Fcal_1^{\rm odd}+(a^2+\m_1\m_2)\, \Fcal_1^{\rm even}\,.
\eq
As in the $N_f=1$ case, generally odd powers of $\lambda$ can not be avoided. However, 
for the special choice $\m_1=-\m_2$ all odd power terms vanish, as can be inferred by 
expanding as well the higher instanton sectors. So that case is very similar to the
pure gauge theory. Note that as in the $N_f=1$ case, $\beta$ dependence in all terms 
of order $\lambda^{-1}$ can be factored out. The $\Z_2$ parity symmetry is broken for 
$\beta\neq 1$ as well.

\subparagraph{$N_f=3$:}

The masses of the flavors are $\vec \m=(\m_1,\m_2,\m_3)$. The $1$-instanton sector reads
\beq
\eqlabel{oneinst3}
\begin{split}
\log Z^{\it inst}(a,\vec \m,\beta^{1/2}\lambda,-\beta^{-1/2}\lambda)|_{q^1}
=&(a^2+\m_1(\m_2+\m_3)+\m_2\m_3)\, \Fcal_1^{\rm odd}\\
&+(a^2(\m_1+\m_2+\m_3)+\m_1\m_2\m_3)\,\, \Fcal_1^{\rm even}\,.
\end{split}
\eq
As in the $N_f=1$ case, generally odd powers of $\lambda$ can not be avoided. However, 
expanding as well the higher instanton sectors shows that in the massless case ($\m_i=0$) 
almost all terms of order $\lambda^{-1}$ drop out, and we have $\Tcal=-\frac a4 + \frac{1}{4}$.

\subparagraph{$N_f=4$:}
The four flavor case with mass vector $\vec \m=(\m_1,\m_2,\m_3,\m_4)$ is very similar to 
the two flavor case. Therefore we will be brief. The main observation is that for choosing 
two pairs of masses with different sign, the odd powers of $\lambda$ drop out, for example 
for the choice $\m_1=-\m_2$ and $\m_3=-\m_4$.

\end{document}